\documentclass[a4paper]{article}
\usepackage{ISCSLP2026}
\usepackage[
    colorlinks=true,
    linkcolor=black,
    citecolor=black,
    urlcolor=black
]{hyperref}
\usepackage{url}
\usepackage{float}
\usepackage{ifthen}
\newboolean{blind}
\setboolean{blind}{false}
\usepackage{siunitx}
\usepackage{cite}
\usepackage{hyperref}
\usepackage{caption}
\usepackage{booktabs}
\usepackage{multirow}
\usepackage{dblfloatfix}
\usepackage{arydshln}
\usepackage{flushend}

\def\RR{{\mathbb R}}

\title{
Separate First, Then Associate:\\A Two-Stage Approach for Real-World Audio-Visual Speech Enhancement
}
\name{
	\ifthenelse{\boolean{blind}}{Anonymous to ISCSLP}
	{
    Tongtao Ling$^1$, Zhenlong Guo$^2$, and Zhong-Qiu Wang$^1$
    \thanks{\noindent This research was supported by National Key Research and Development Program of China (Grant No. 2025YFF0518003). \textit{Corresponding author: Zhong-Qiu Wang}.}
    }
}

\address{
  \ifthenelse{\boolean{blind}}{Anonymous to ISCSLP}
    {
  	$^1$ Southern University of Science and Technology, Shenzhen, China \\
    $^2$ Xiaomi Inc., Beijing, China
    }
}

\email{
	\ifthenelse{\boolean{blind}}{Anonymous to ISCSLP}
	{lingtt2025@mail.sustech.edu.cn, wang.zhongqiu41@gmail.com}
}

\begin{document}

\maketitle
\begin{abstract}
Audio-visual speech enhancement (AVSE) aims at extracting target speech from multi-speaker mixtures by exploiting visual cues. Although recent studies have reported strong performance on simulated datasets, the performance, however, often drops dramatically when they are applied to real-world audio-visual recordings.
To bridge this gap, the Real-World AVSE Challenge held in the ISCSLP 2026 conference calls for participants to design a practical solution for AVSE under real-world conditions, where speaker overlap, acoustic interferences, room reverberation and visual degradations naturally co-exist.
In our submission to the challenge, we propose a decoupled \textit{separation-then-association}
approach.
It consists of two stages: a separation stage in which a trained, audio-only model (i.e., not using visual cues) is used to separate input multi-speaker mixture to individual speaker signals, followed by an association stage, where an audio-visual CLIP model is used to 
identify the separated speech signal with the highest similarity with the target speaker's facial video via cross-modal similarity matching.
Evaluation results on the challenge dataset show the effectiveness of our proposed approach.
\end{abstract}

\section{Introduction}

Audio-only speech separation systems have made substantial progress in recent years, since the speaker permutation ambiguity problem was successfully addressed by optimizing permutation invariant objectives \cite{Hershey2016,yu2017permutation,wang2023tf,ling2026mc}.
Although individual speech sources can be successfully separated, the system does not inherently know which output corresponds to the target speaker.
Visual cues provide a natural solution to this problem.
The motion of a speaker's face, particularly around the mouth region, is strongly correlated with the corresponding speech signal.
Audio-visual speech enhancement (AVSE) therefore exploits visual cues to identify or extract the speech of the target speaker from a multi-speaker mixture \cite{Ephrat2018,Gu2020,Michelsanti2021,tao2025audio,pan2022selective,pan2023scenario,mu2024separate,kalkhorani2025av,ling2026tgtse,ling26_chime}.
Most existing AVSE systems incorporate visual representations directly into the speech separation network.
In such systems, visual features are temporally aligned with acoustic representations and used to condition the separation process.

The Real-World AVSE Challenge \cite{organizers2026iscslp2026realworldaudiovisual}
further extends this problem to realistic recording conditions.
In contrast to benchmarks constructed through synthetic mixing, the Track $1$ of the challenge contains naturally-recorded speech mixtures with realistic reverberation, ambient noises, and speech overlap.
In addition, Track $2$ is designed to deal with the case where visual cues suffer from degradations such as low resolution, occlusion, missing or frozen frames, and time-synchronization issues.
These conditions make tightly-coupled audio-visual separation more difficult because unreliable visual features can directly interfere with the speech separation process \cite{yang2026multi,li2025momuse,cheng2025multi}.

In this paper, we explore an alternative strategy which decouples speech separation with audio-visual speaker association.
Instead of using visual cues to condition and guide the separator, we first perform multi-speaker separation using strong audio-only models such as TF-GridNet \cite{wang2023tf}.
We then employ a pre-trained audio-visual contrastive model (i.e., AV-CLIP \cite{iashin2024synchformer}) to measure the correspondence between each separated signal and the target speaker video.
The separated signal with higher audio-visual similarity is selected as estimated target speech.

Our approach has several advantages.
First, speech separation is performed entirely in the audio domain, preventing degraded visual cues from directly affecting the separation process.
Second, the audio-visual model only needs to solve a relatively simple speaker association problem rather than the more difficult speech separation problem.
Third, the separation and association modules can be trained independently using substantially different datasets, making it possible to exploit large-scale audio-only speech separation data together with audio-visual speech data which is of limited availability.

It should be noted that there are early studies in target speaker extraction, which first perform multi-speaker separation and then conduct speaker verification based on extracted speaker embeddings to select the target speaker \cite{rao2019target,malek2022target,vzmolikova2019speakerbeam}.
A similar separation-and-selection paradigm is investigated recently in brain-assisted speech enhancement \cite{xu2026analysis}, where auditory attention decoding is used to identify the attended speech after multi-speaker separation.
Differently, our study extends this approach to AVSE, a different task, and we leverage audio-visual contrastive learning to perform target speaker selection.

\section{System Description}

\begin{figure*}
    \centering
    \includegraphics[width=1.0\linewidth]{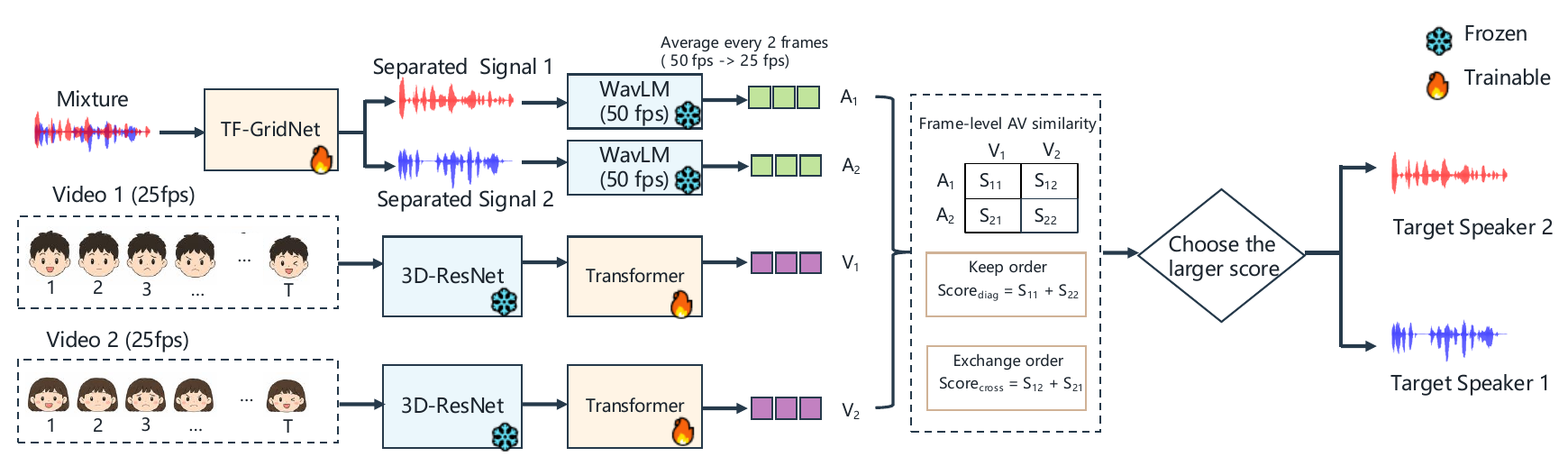}
    \vspace{-0.6cm}
    \caption{Overview of proposed separation-then-association approach for AVSE.}
    \vspace{-0.6cm}
    \label{fig:model}
\end{figure*}

Figure \ref{fig:model} shows our submitted system, consisting of three components: a speech separator, an audio-visual CLIP model, and a cross-modality, audio-visual speaker association module.

\subsection{Speech Separator} 

We employ TF-GridNet \cite{wang2023tf} for speech separation.
It exhibits strong performance in multiple recent speech separation benchmarks.
Given a two-speaker mixture signal $y$, the separator estimates two candidate speech signals, $\hat{s}_1$ and $\hat{s}_2$, solely based on the audio mixture without using any visual cues.
Since speech separation is inherently permutation-invariant, the ordering of the separated outputs is not associated with that of the visible speakers.
In other words, the separator determines \emph{what speech signals are present} in the mixture but does not determine \emph{which separated signal belongs to which visible speaker}. We resolve this ambiguity in the subsequent cross-modal speaker association stage, where a trained AV-CLIP model is used to measure the temporal correspondence between each separated speech candidate and each visible speaker.

\subsection{Frame-level AV-CLIP}

Unlike segment-level contrastive learning \cite{iashin2024synchformer}, the objective of using AV-CLIP in our task is to align visual representations with speech representations through frame-level contrastive learning \cite{radford2021learning}.
After that, we evaluate the correspondence between each separated signal and the target video, and determines which candidate belongs to the target speaker. 
In detail, given a synchronized audio-visual pair, we first employ a pre-trained self-supervised speech encoder (e.g., WavLM \cite{chen2022wavlm}) to extract frame-level acoustic representations:
\begin{equation}
\mathbf{A} = [a_1,a_2,\dots,a_{T_a}]\in \RR^{T_a\times D},
\end{equation}
where $a_t\in\mathbb{R}^{D}$ is a $D$-dimensional embedding corresponding to frame $t$ of audio, and $T_a$ is the total number of audio frames.
Meanwhile, we use a pre-trained visual encoder (e.g., 3D-ResNet18 \cite{afouras2018deep}) to extract frame-level visual representations.
Since each visual frame is independently encoded, we feed the sequence of frame embeddings to a trainable Transformer encoder \cite{vaswani2017attention} to capture temporal dependencies among adjacent lip movements, resulting in:
\begin{equation}
\mathbf{V}
=
[v_1,v_2,\dots,v_{T_v}]\in \RR^{T_v\times D},
\end{equation}
where $v_t \in \mathbb{R}^{D}$ is the visual embedding at video frame $t$, and $T_v$ the total number of visual frames.
Since WavLM operates at $50$ frames per second (fps) while the video stream is sampled at $25$ fps, the acoustic embeddings have approximately twice the temporal resolution of the visual embeddings (i.e., $2T_v \approx T_a$).
To align the two modalities, we average every two consecutive audio frames, resulting in temporally aligned audio and visual embeddings with the same frame rate. That is, $T \approx T_v \approx \frac{1}{2}T_a$.

Based on the time-aligned embeddings, i.e., $\widetilde{\mathbf{A}}\in\mathbb{R}^{T\times D}$ and $\widetilde{\mathbf{V}}\in\mathbb{R}^{T\times D}$,  we follow an InfoNCE loss \cite{radford2021learning} to perform audio-visual contrastive learning at the frame level.
Audio and visual embeddings from the same temporal frame are treated as positive pairs, while those from different frames are treated as negative pairs.
This objective encourages temporally-corresponding audio and visual embeddings to be close to each other in the shared embedding space while pushing away non-corresponding embeddings, thereby enabling the model to learn fine-grained correspondence between speech and lip movements.
In detail, the similarity matrix between visual and acoustic embeddings is computed as
\begin{equation}
\mathbf{S}
=
\frac{1}{\tau}\cdot\widetilde{\mathbf{V}}
\widetilde{\mathbf{A}}^{\top}
\in \mathbb{R}^{T \times T},
\end{equation}
where $\tau$ is a learnable temperature parameter.
For each visual frame, the synchronized audio frame is regarded as the positive sample, while all remaining frames serve as negative samples. Consequently, the visual-to-audio contrastive loss is defined as
\begin{equation}
\mathcal{L}_{\text{v2a}}
=
-
\frac{1}{T}
\sum_{i=1}^{T}
\log
\frac
{\exp(S_{ii})}
{\sum_{j=1}^{T}\exp(S_{ij})},
\end{equation}
and, similarly, the audio-to-visual objective is formulated as
\begin{equation}
\mathcal{L}_{\text{a2v}}
=
-
\frac{1}{T}
\sum_{i=1}^{T}
\log
\frac
{\exp(S_{ii})}
{\sum_{j=1}^{T}\exp(S_{ji})}.
\end{equation}
The overall temporal alignment loss is
$\mathcal{L}_{\text{av-clip}}
=
\mathcal{L}_{\text{v2a}}
+
\mathcal{L}_{\text{a2v}}$.

\subsection{Cross-Modal Speaker Association}

After speech separation, the two separated signals are permutation-invariant (i.e., the separation model itself does not determine which output corresponds to which visible speaker).
We therefore employ the trained AV-CLIP model to establish the correspondence between the separated signals and the visible speakers based on their temporal audio-visual consistency.

Given two separated speech candidates $A_1$ and $A_2$ and two synchronized target videos $V_1$ and $V_2$, we first extract their frame-level audio and visual embeddings using the audio and visual encoders of frame-level AV-CLIP.
As described in the previous section, the audio embeddings are temporally down-sampled to approximately $25$ fps to match the frame rate of the visual embeddings.
The resulting embeddings are then $\ell_2$-normalized along the embedding dimension.
In detail, for each audio-video pair $(A_i,V_j)$ between the audio of speaker $i$ and video of speaker $j$, we compute a frame-level cosine similarity between temporally-corresponding audio and visual embeddings.
The overall audio-visual matching score is obtained by averaging the similarities over all valid frames:
\begin{equation}
S_{i,j}
=
\frac{1}{T}
\sum_{t=1}^{T}
\frac{
\mathbf{a}_{i,t}^{\top}\mathbf{v}_{j,t}
}{
|| \mathbf{a}_{i,t}||_2
|| \mathbf{v}_{j,t}||_2
},
\end{equation}
where $\mathbf{a}_{i,t}$ denotes the audio embedding of the $i$-th separated signal at frame $t$, $\mathbf{v}_{j,t}$ the visual embedding of the $j$-th speaker, and $T$ the number of temporally-aligned frames.
A higher $S_{i,j}$ indicates stronger temporal correspondence between the separated speech and the lip movements.
Since there are two speakers, only two possible one-to-one assignments need to be considered. Their association scores are respectively computed as
\begin{equation}
S_{\mathrm{diag}} = S_{1,1}+S_{2,2},
\qquad
S_{\mathrm{cross}} = S_{1,2}+S_{2,1},
\end{equation}
and the final permutation is selected according to
\begin{equation}
\pi^{*}=
\begin{cases}
(1,2), & S_{\mathrm{diag}} \geq S_{\mathrm{cross}};\\
(2,1), & \text{otherwise}.
\end{cases}
\end{equation}
Accordingly, the separated signals are reordered such that each output is associated with its corresponding visible speaker. In this way, AV-CLIP resolves the permutation ambiguity of speech separation without requiring a speaker enrollment utterance or an explicit speaker identity model.

\section{Experimental Setup}


In the Real-World AVSE Challenge, all audio signals are provided as monaural recordings sampled at $16$ kHz, and each recording contains two speakers.
Since no dedicated training set is provided by the challenge and the participants are allowed to use any public datasets for model training, we first train a speech separation model on the Libri2Mix dataset \cite{cosentino2020librimix}, and subsequently fine-tune it using the real-recorded development data provided by the challenge \cite{yang2026identityfaithfulaudiovisualtargetspeaker}.
As shown in Table \ref{tab:statistic}, the development and test sets consist of two types of mixture signals: \textit{mix} and \textit{remix}.
The \textit{mix} data correspond to naturally recorded real-world mixtures without clean reference signals and are therefore evaluated only using non-intrusive metrics.
In contrast, the \textit{remix} data are constructed by mixing two clean, real-recorded single-speaker recordings, providing clean references for reference-based evaluation.
We hence leverage the \textit{remix} data with available clean references for supervised fine-tuning of the separation model.
This two-stage training strategy allows the model to first learn general speech separation capabilities from Libri2Mix, and subsequently adapt to the acoustic characteristics of the challenge recordings, thereby reducing domain mismatches between simulated and real-world conditions. 


We employ TF-GridNet \cite{wang2023tf} to separate each
mixture to two candidate signals. Each training segment is $4$-second long. Following the hyper-parameters in TF-GridNet \cite{wang2023tf}, we set $B=6$, $D=128$, $H=256$, $I=1$, $J=1$, $E=8$, and $L=4$.
For short-time Fourier transform (STFT), we use $16$ ms window, $8$ ms hop, and a square-root Hann analysis window.
The model is trained using the scale-invariant signal-to-noise ratio (SI-SNR) \cite{le2019sdr} objective with permutation invariant training (PIT) \cite{yu2017permutation}.
We train TF-GridNet for up to $100$ epochs, using the Adam optimizer \cite{kingma2014adam} with a batch size of $2$.
The initial learning rate is set to $5\times10^{-4}$ and is reduced by half if the validation loss does not improve for two consecutive epochs. 

To train AV-CLIP, we use the LRS$3$ dataset \cite{afouras2018lrs3}.
For the audio branch, we employ WavLM-Large
as the audio encoder and apply a linear projection layer to map its $1,024$-dimensional embeddings to $512$-dimensional.
For the visual branch, we use 3D-ResNet18 as the visual encoder, followed by a Transformer encoder for temporal modeling.
During AV-CLIP training, the parameters of both the pretrained WavLM-Large and 3D-ResNet18 encoders are frozen, while the projection layer and Transformer encoder are optimized for learning audio-visual correspondence.
The Transformer consists of $6$ layers, each with a hidden dimension of $512$, a feed-forward dimension of $2,048$, and $8$ self-attention heads, resulting in around $19.4$ million trainable parameters. We optimize the model using the AdamW optimizer \cite{loshchilov2017decoupled} with an initial learning rate of $10^{-4}$ and a cosine learning rate schedule. Specifically, the learning rate is linearly warmed up in the first $10\%$ of the training epochs and then decayed according to a cosine schedule for the remaining epochs.
The model is trained for $100$ epochs with a batch size of $128$.
To better adapt the learned audio-visual embeddings to the target domain, we further fine-tune AV-CLIP on the development \textit{remix} subset, which provides synchronized audio-visual samples that more closely match the evaluation conditions.

\begin{table}[t]
    \centering
    \footnotesize
    \sisetup{table-format=2.1,round-mode=places,round-precision=1,table-number-alignment=center,detect-weight=true,detect-inline-weight=math}
    \caption{Data statistics of Real-World AVSE Challenge.}
    \vspace{-0.2cm}
    \begin{tabular}{
    c
    c
    S[table-format=4,round-precision=0] 
    S[table-format=4,round-precision=0] 
    }
    \toprule
    Split & Scene & {Track 1} & {Track 2} \\
    \midrule
    dev	& mix &   1242 & 1527 \\
    dev	& remix	&  900 & 1098 \\
    test & mix &  2472 & 2820 \\
    test & remix & 1785 & 2121 \\
    \bottomrule
    \end{tabular}
    \vspace{-0.6cm}
    \label{tab:statistic}
\end{table}

\begin{table*}[t]
    \centering
    \caption{Evaluation results on Track 1.}
    \vspace{-0.2cm}
    \sisetup{table-format=2.1,round-mode=places,round-precision=1,table-number-alignment=center,detect-weight=true,detect-inline-weight=math}
    \resizebox{\textwidth}{!}{
    \begin{tabular}{
    l|
    c
    S[table-format=2.1,round-precision=1] 
    S[table-format=1.2,round-precision=2] 
    S[table-format=1.3,round-precision=3] 
    S[table-format=1.3,round-precision=3] 
    S[table-format=1.3,round-precision=3] 
    S[table-format=1.3,round-precision=3] 
    S[table-format=1.3,round-precision=3] 
    S[table-format=1.3,round-precision=3] 
    S[table-format=3.1,round-precision=1] 
    S[table-format=1.3,round-precision=3] 
    }
    \toprule
    Model & Scene & {SI-SDR (dB)$\uparrow$} & {PESQ$\uparrow$} & {STOI$\uparrow$} & {UTMOS$\uparrow$} & {DNS-P808$\uparrow$} & {DNS-SIG$\uparrow$} & {DNS-BAK$\uparrow$} & {DNS-OVRL$\uparrow$} & {CER ($\%$)$\downarrow$} & {SPK-SIM$\uparrow$} \\
    \midrule
    Baseline (AV-ConvTasNet) & mix & {--} & {--} & {--} & 0.7295 & 2.1839 & 1.7935 & \bfseries 2.8431 & 1.4668 & 105.90 & 0.3336 \\
    Baseline (AV-ConvTasNet) & remix & -5.9252 & 1.1373 & 0.3036 & 0.9263 & 2.1597 & 1.7766 & 2.5115 & 1.4200 & 97.89 & 0.3197 \\
    Baseline (AV-ConvTasNet) & both & -5.9252 & 1.1373 & 0.3036 & 0.8120 & 2.1737 & 1.7864 & 2.7041 & 1.4472 & 102.54 & 0.3278 \\
    [0.5ex]\hdashline\noalign{\vskip 0.5ex}
    Separation-then-association  & mix & {--} & {--} & {--} & 1.9433 & \bfseries 2.8265 & \bfseries 2.2407 & 2.3450 & \bfseries 1.7854 & 18.62 & 0.7203 \\
    Separation-then-association & remix & \bfseries 10.4156 & \bfseries 2.6511 & \bfseries 0.8232	& 2.0725 & 2.7671 & 2.1506 & 2.3209 & 1.7487 & \bfseries 13.24 & \bfseries 0.7592 \\
    Separation-then-association & both & \bfseries 10.4156 & \bfseries 2.6511 & \bfseries 0.8232 & 1.9975 & 2.8016 & 2.2029 & 2.3349 & 1.7700 & 16.37 & 0.7366 \\
    \quad + Dynamic mixing & mix & {--} & {--} & {--} & 1.9397 & 2.7949 & 2.2327 & 2.2856 & 1.7476 & 20.74 & 0.7164 \\
    \quad + Dynamic mixing & remix & 10.0321 & 2.6269 & 0.8211 & \bfseries 2.0945 & 2.7541 & 2.1267 & 2.2072 & 1.7062 & 13.44 & 0.7531  \\
    \quad + Dynamic mixing & both & 10.0321 & 2.6269 & 0.8211 & 2.0046 & 2.7778 & 2.1883 & 2.2528 & 1.7302 & 17.68 & 0.7318  \\
    \bottomrule
    \end{tabular}
    }
    \vspace{-0.3cm}
    \label{tab:track1}
\end{table*}

\begin{table*}[t]
    \centering
    \sisetup{table-format=2.1,round-mode=places,round-precision=1,table-number-alignment=center,detect-weight=true,detect-inline-weight=math}
    \caption{Evaluation results on Track 2.}
    \vspace{-0.2cm}
    \resizebox{\textwidth}{!}{
    \begin{tabular}{
    l|
    c
    S[table-format=2.1,round-precision=1] 
    S[table-format=1.2,round-precision=2] 
    S[table-format=1.3,round-precision=3] 
    S[table-format=1.3,round-precision=3] 
    S[table-format=1.3,round-precision=3] 
    S[table-format=1.3,round-precision=3] 
    S[table-format=1.3,round-precision=3] 
    S[table-format=1.3,round-precision=3] 
    S[table-format=3.1,round-precision=1] 
    S[table-format=1.3,round-precision=3] 
    }
    \toprule
    Model & Scene & {SI-SDR (dB)$\uparrow$} & {PESQ$\uparrow$} & {STOI$\uparrow$} & {UTMOS$\uparrow$} & {DNS-P808$\uparrow$} & {DNS-SIG$\uparrow$} & {DNS-BAK$\uparrow$} & {DNS-OVRL$\uparrow$} & {CER ($\%$)$\downarrow$} & {SPK-SIM$\uparrow$} \\
    \midrule
    Baseline (AV-ConvTasNet) & mix & {--} & {--} & {--} & 1.0965 & 2.3423 & 1.3718 & 1.3594 & 1.1658 & 109.36 & 0.4007 \\
    Baseline (AV-ConvTasNet) & remix & -1.6892 &  1.3036 & 0.5021 & 1.1039 & 	2.3067 & 1.5282	 & 1.5371 & 1.2621 & 99.76 & 0.3892 \\
    Baseline (AV-ConvTasNet) & both & -1.6892 & 1.3036	& 0.5021 & 1.0996 &	2.3271 & 1.4390 & 1.4357 & 1.2071 & 105.24 & 0.3957 \\
    [0.5ex]\hdashline\noalign{\vskip 0.5ex}
    Separation-then-association  & mix & {--} & {--} & {--} & 1.9653 & \bfseries 2.8053 & \bfseries 2.1966 & \bfseries 2.2974 & \bfseries 1.7545 & 23.06 & 0.7104 \\
    Separation-then-association & remix & \bfseries 9.5414 & 2.5542 & 0.8045 & \bfseries 2.0908 & 2.7285 & 2.0892 & 2.2605 & 1.7047 & 18.43 & 0.7462 \\
    Separation-then-association & both & \bfseries 9.5414	& 2.5542 & 0.8045 & 2.0192 & 2.7723 &2.1505 & 2.2816 & 1.7331 & 21.08	& 0.7258 \\
    \quad + Dynamic mixing  & mix & {--} & {--} & {--}  & 1.8086 &	2.7300	& 1.8476 &	1.8886&	1.5145	& 22.52	&0.7526 \\
    \quad + Dynamic mixing & remix & 9.2707	& \bfseries 2.5935 & \bfseries 0.8047 &1.9619&	2.6531	& 1.8387 & 1.9472	& 1.5355 & \bfseries 17.97 & \bfseries 0.7775 \\
    \quad + Dynamic mixing & both & 9.2707 &	\bfseries 2.5935	& \bfseries 0.8047 & 1.8744 &2.6970	& 1.8438 & 1.9138	& 1.5235 & 20.57 & 0.7633 \\
    \bottomrule
    \end{tabular}
    }
        \vspace{-0.3cm}
    \label{tab:track2}
\end{table*}

\begin{table*}[!ht]
\centering
\footnotesize
\sisetup{table-format=2.1,round-mode=places,round-precision=1,table-number-alignment=center,detect-weight=true,detect-inline-weight=math}
\caption{Comparison with other participating systems on Track 1 and 2.}
\vspace{-0.2cm}
\label{tab:challenge_results}
\resizebox{\textwidth}{!}{
    \begin{tabular}{
    l
    c
    S[table-format=1.3,round-precision=1] 
    S[table-format=1.3,round-precision=2] 
    S[table-format=1.3,round-precision=2] 
    S[table-format=1.3,round-precision=3] 
    S[table-format=1.3,round-precision=3] 
    S[table-format=1.3,round-precision=1] 
    S[table-format=1.3,round-precision=3] 
    S[table-format=1.3,round-precision=2] 
    }
\toprule
{Track} &
{System} &
{SI-SDR (dB)$\uparrow$} &
{PESQ$\uparrow$} &
{STOI$\uparrow$} &
{UTMOS$\uparrow$} &
{DNS-OVRL$\uparrow$} &
{CER(\%)$\downarrow$} &
{SPK-SIM$\uparrow$} &
{OVRL$\downarrow$} \\
\midrule

\multirow{6}{*}{Track 1}
& Baseline (AV-ConvTasNet)
& -5.9 & 1.137 & 0.304 & 0.812 & 1.447 & 102.5 & 0.328 & 16.14  \\
\cmidrule(l){2-10}

& audioman
& 10.70 & 2.933 & 0.841 & 2.074 & 1.832 & 12.3 & 0.749 & 2.86 \\

& AITD
& 12.72 & 3.000 & 0.852
& 2.100 & 1.697 & 14.5 & 0.769 & 2.86 \\

& twilight
& 9.16 & 2.775 & 0.825 & 2.145 & 1.944
& 13.5 & 0.742 & 3.57 \\

& YiJiaHe
& 10.23 & 2.717 & 0.817
& 2.765 & 2.022 & 17.1 & 0.726 & 4.00 \\

& \textbf{ Separation-then-association (Ours)}
& 10.42 & 2.651 & 0.823 & 1.998 & 1.770 & 16.4 & 0.737 & 5.43 \\
\midrule

\multirow{6}{*}{Track 2}
& Baseline (AV-ConvTasNet)
& -1.69 & 1.304 & 0.502 & 1.100 & 1.207 & 105.2 & 0.396 & 11.29  \\
\cmidrule(l){2-10}

& AITD
& 12.26 & 2.939 & 0.844
& 2.129 & 1.670 & 15.3 & 0.764 & 2.43 \\

& audioman
& 10.22 & 2.888 & 0.830 & 2.094 & 1.760 & 14.8 & 0.744 & 2.86 \\

& YiJiaHe
& 8.93 & 2.615 & 0.789
& 2.766 & 2.012 & 22.0 & 0.707 & 4.14 \\

& twilight
& 7.05 & 2.561 & 0.776 & 2.170 & 1.896 & 22.0 & 0.712 & 5.00 \\

& \textbf{ Separation-then-association (Ours)}
& 9.54 & 2.554 & 0.805 & 2.019 & 1.733 & 21.1 & 0.726 & 5.29 \\

\bottomrule
\end{tabular}
}
    \vspace{-0.5cm}
\end{table*}


We evaluate the system using both intrusive (reference-based) and non-intrusive (reference-free) metrics. For intrusive evaluation, we employ scale-invariant signal-to-distortion ratio (SI-SDR) \cite{le2019sdr}, perceptual evaluation of speech quality (PESQ) \cite{rix2001perceptual} and short-time objective intelligibility (STOI) \cite{taal2010short} to respectively measure speech separation quality, perceptual quality, and intelligibility. Since the real-recorded \textit{mix} data do not provide clean reference signals, these intrusive metrics are only computed on the \textit{remix} subset. 
For non-intrusive evaluation, we report UTMOS \cite{baba2024t05} and DNSMOS \cite{reddy2022dnsmos} for perceptual speech quality assessment, character error rate (CER) computed using Fun-ASR-Nano \cite{an2025fun} to evaluate speech recognition performance, and speaker similarity (SPK-SIM) computed using WeSpeaker-ResNet34 \cite{wang2023wespeaker} to assess the preservation of target-speaker characteristics. These reference-free metrics are evaluated on both the \textit{mix} and \textit{remix} subsets.


We compare our proposed system with the challenge baseline, AV-ConvTasNet \cite{wu2019time}. In addition, we evaluate a variant of our system, where the speech separator is further trained with dynamically generated mixtures (i.e., dynamic mixing \cite{zeghidour2021wavesplit}) to investigate whether additional simulated training mixtures can improve its generalizability to real-world recordings.

\section{Evaluation Results}

Table \ref{tab:track1} and \ref{tab:track2} respectively report results on Track 1 and 2 of the challenge.
We compare our separation-then-association approach with the official AV-ConvTasNet baseline under three evaluation settings, including \textit{mix}, \textit{remix} and \textit{both}.
Since the \textit{mix} data are naturally-recorded mixtures without clean reference signals, intrusive metrics (such as SI-SDR, PESQ and STOI) are only reported for the \textit{remix} data.

Compared with the AV-ConvTasNet baseline, our separation-then-association approach obtains substantially better performance in both tracks.
On Track 1, the SI-SDR on the \textit{remix} set improves from $-5.9$ to $10.4$ dB, amounting to an absolute improvement of $16.3$ dB.
Meanwhile, PESQ and STOI increase from $1.14$ and $0.304$ to $2.65$ and $0.823$, respectively.
Significant improvements are also observed for reference-free metrics.
On the combined \textit{both} set, UTMOS improves from $0.812$ to $1.998$, while CER is reduced from $102.5\%$ to $16.4\%$ and SPK-SIM increases from $0.328$ to $0.737$.
These results indicate that the proposed approach not only achieves better signal reconstruction but also better preserves speech intelligibility and target-speaker characteristics.

A similar trend is observed on Track 2.
Compared with the AV-ConvTasNet baseline, our system improves SI-SDR from $-1.7$ to $9.5$ dB on the \textit{remix} set, and increases PESQ and STOI from $1.30$ and $0.502$ to $2.55$ and $0.805$, respectively.
On the combined set, CER is reduced from $105.2\%$ to $21.1\%$, while SPK-SIM improves from $0.396$ to $0.726$.
The consistent improvements in both tracks demonstrate the effectiveness and robustness of our proposed approach for real-world AVSE.

We further investigate whether dynamically-generated speech mixtures (i.e., dynamic mixing \cite{zeghidour2021wavesplit}) can improve the generalizability of the separation model.
The results show different behaviors on the two tracks.
On Track 1, dynamic mixing  provides no consistent improvement: SI-SDR slightly decreases from $10.4$ to $10.0$ dB, accompanied by small degradations in PESQ and STOI.
The reference-free metrics exhibit a similar trend, suggesting that dynamically-generated mixtures still exhibit domain mismatches with real-world recordings.

On Track 2, however, dynamic mixing improves target-speaker preservation despite some degradation in perceptual speech quality.
In particular, SPK-SIM increases from $0.710$ to $0.753$ on \textit{mix}, from $0.746$ to $0.778$ on \textit{remix}, and from $0.726$ to $0.763$ on the combined set.
CER is also slightly reduced from $23.1\%$ to $22.5\%$ on \textit{mix} and from $18.4\%$ to $18.0\%$ on \textit{remix}.
In contrast, SI-SDR decreases slightly from $9.5$ to $9.3$ dB, and several DNSMOS-related metrics also decrease.
This indicates a trade-off between perceptual speech quality and target-speaker preservation: dynamic mixing may improve the ability of the model to retain target-speaker characteristics without necessarily improving conventional signal-level metrics.

Table~\ref{tab:challenge_results} compares our system with other participating systems.
On Track 1, our system achieves $10.4$ dB SI-SDR, which is competitive with audioman and YiJiaHe, although lower than the best-performing AITD system ($12.7$ dB). Similar trends are observed for PESQ and STOI. In terms of target-speaker preservation, our system obtains a SPK-SIM of $0.737$, outperforming YiJiaHe and remaining close to twilight and audioman. On Track 2, our method achieves $9.5$ dB SI-SDR, outperforming YiJiaHe and twilight, while approaching audioman ($10.2$ dB). Our STOI of $0.81$ also exceeds those of YiJiaHe and twilight. For SPK-SIM, our system achieves $0.726$, higher than YiJiaHe and twilight, although AITD and audioman obtain better results. These results suggest that the proposed framework achieves competitive separation performance despite its relatively simple decoupled architecture. However, our system shows competitive performance across both tracks without being specifically optimized for a single evaluation metric. 


\section{Conclusions}


We have proposed a simple yet effective separation-then-association approach for real-world AVSE.
Instead of directly extracting target speech conditioned on lip embeddings, our framework decomposes AVSE into two stages: speech separation and cross-modal speaker association.
Evaluation results on the challenge dataset demonstrate that our approach substantially outperforms the challenge baseline across a wide range of evaluation metrics, confirming the effectiveness of the proposed framework for challenging real-world audio-visual scenarios.



\bibliographystyle{IEEEtran}

\bibliography{mybib}


\end{document}